\documentstyle[prl,aps,epsf]{revtex}
\begin{document}
\twocolumn[\hsize\textwidth\columnwidth\hsize \csname @twocolumnfalse\endcsname

\title{
Magnetoresistance, Micromagnetism, and Domain Wall Scattering \\
in Epitaxial hcp Co Films
}

\author{U. R\"{u}diger$^1$, J. Yu$^1$, L. Thomas$^2$, S. S. P. Parkin$^2$,
and  A. D. Kent$^{1,*}$}

\address{
$^1$ Department of Physics, New York University,
4 Washington Place, New York, New York 10003, USA}

\address{
$^2$ IBM Research Division, Almaden Research Center,
San Jose, California  95120, USA}

\date{July 31, 1998}

\maketitle

\begin{abstract}
Large negative magnetoresistance (MR) observed in
transport measurements of hcp Co films with stripe domains were 
recently reported and interpreted in terms of a novel domain wall (DW) 
scattering mechanism. Here detailed MR measurements, magnetic force 
microscopy, and micromagnetic calculations are combined to 
elucidate the origin of MR in this material. The large negative room 
temperature MR reported previously is shown to be due to ferromagnetic 
resistivity anisotropy. Measurements of the resistivity for currents
parallel (CIW) and perpendicular to DWs (CPW) have been conducted as a function
of temperature. Low temperature results show that any intrinsic effect of DWs 
scattering on MR of this material is very small compared to
the anisotropic MR.
\end{abstract}
\label{firstpage}
]
\narrowtext
The effect of magnetic domain walls (DWs) on the transport properties of
thin films and nanostructures is a topic of great current interest.
Recent experimental research has extended early studies of iron
single crystals \cite{Coleman,Berger} to novel nanofabricated thin film 
structures of 3d transition metals \cite{Viret,Gregg,Ulrich} and 
transition metal alloys \cite{Shinjo,Dafine}. This topic
has been approached from a number of viewpoints.
In nanowires an experimental goal
has been to use magnetoresistance (MR) to investigate DW nucleation and 
dynamics in search of evidence for macroscopic quantum phenomena. 
Conductance fluctuations and MR hysteresis observed at low temperature
in nanowires of Ni, Fe and Co \cite{Hong,Otani} have 
stimulated new theoretical work on the 
effect of DWs on quantum transport in mesoscopic ferromagnetic
conductors \cite{Tatara,Geller}. In thin films and
microstructures with stripe domains, experiments have focused on
understanding the basic mechanisms of DW scattering of conduction
electrons.  Specifically, large negative MR observed at room temperature
in hcp Co thin films with stripe domains were recently
reported and interpreted in terms of a giant DW scattering contribution 
to the resistivity \cite{Gregg}.  Independently, and to understand this result,
a new mechanism of DW scattering was
proposed which invokes the two channel model of conduction in 
ferromagnets and spin dependent electron scattering -- a starting point for 
understanding the phenomena of giant MR (GMR) \cite{Levy}. 
Within this model DWs 
increase resistivity because they mix the minority and majority spin 
channels and thus partially eliminate the short-circuit provided by the 
lower resistivity spin channel in the magnetically homogeneous 
ferromagnet. 

Here we present a new physical interpretation of the MR of hcp Co films with
stripe domains which is based on both new experimental results and micromagnetic
modeling.  We have conducted experiments on samples of systematically varied 
magnetic structure and DW density and as a function of the angle of the 
transport current with respect to DWs. The role of conventional sources 
of MR in ferromagnetic metals on the interpretation of such experiments
is discussed in detail. MR measurements, magnetic force 
microscopy (MFM) imaging in conjunction with micromagnetic simulations 
show that the large negative MR observed at room temperature in hcp Co films for 
fields applied parallel to the the easy magnetic axis is due to a 
conventional anisotropic transport effect in ferromagnetic metals, 
not large DW scattering effects. 

Epitaxial (0001) oriented hcp Co films of 55 nm, 70 nm, 145 nm,
and 185 nm thicknesses have been studied. The films were grown on a-axis
($11{\bar 2}$0) sapphire substrates using e-beam evaporation techniques
under UHV conditions. First, at a temperature of 680 K a 10 nm thick
(0001) Ru seed layer was deposited followed by a (0001) Co layer.
The Co layer was protected against corrosion by a 5 nm thick Ru capping
layer. X-ray $\theta/2\theta$-scans 
indicate c-axis orientation of the Ru and Co layers. Off-axis x-ray pole 
figures show that the films are also oriented
in-plane with respect to the sapphire substrate. The films were 
patterned using projection optical lithography and ion milling in order to 
produce microstructures of well defined geometry for MR studies. A residual 
resistivity of $\rho =0.16 \: \mu\Omega$cm and the residual resistivity
ratio of 19 for a 185 nm thick 5 $\mu$m linewidth Co wire confirm the high
crystalline quality of the films. 

These films have a strong uniaxial anisotropy with the magnetic easy 
axis perpendicular to the film plane \cite{Hehn}. The competition between 
magnetostatic, exchange, and magnetocrystalline energies leads to a stripe 
domain configurations in which the domain size depends on the sample 
thickness and the domain configurations depend on the sample magnetic 
history. Fig. 1 shows MFM images of a 70 nm thick 5 $\mu$m linewidth Co 
wire in 
zero magnetic field. These MFM images, taken with a vertically magnetized 
magnetic tip, highlight the out-of-plane component of the wire 
magnetization. Images are shown after magnetic saturation: a) perpendicular 
to the film plane, b) in-plane and transverse to the wire axis, 
and c) in-plane 
and along the wire axis. As seen in Fig. 1,  an 
in-plane applied field can be employed to align DWs in stripes
\cite{Kooy}. Fig. 1b and 1c show 
that DWs can be oriented parallel or perpendicular to the long axis of the
wire and thus the applied current, denoted as current-in-wall (CIW) 
and current-perpendicular-to-wall (CPW) geometries, respectively \cite{Levy} 
(as shown in the drawing in Fig. 1).

\begin{figure}[bt]
\epsfxsize=2.95in
\vspace{8 cm}
\caption{MFM images in zero applied field of a of 5 $\mu$m linewidth
70 nm thick Co wire after (a) perpendicular, (b) transverse, and 
(c) longitudinal magnetic saturation. The model shows the orientation
of stripe and flux closure caps with respect to the current for (b)
CPW and (c) CIW geometries.}
\label{fig6}
\end{figure}

Modeling of the film micromagnetic structure is essential to
understand the MR results. An important parameter 
for stripe domain materials is the ratio of is anisotropy to 
demagnetization energy, known as the quality factor $Q$, given by 
$Q=K/2\pi M^2_s$ \cite{Landau,Kittel,Hubert}.  For small $Q$ ($Q<<1$), 
the magnetostatic energy dominates the 
anisotropy energy.  In this limit, it is energetically favorable to
maintain flux closure at the film boundaries via the formation of closure
domains (with magnetization parallel to the film surface) 
at the top and bottom film surfaces. 
In the limit of large $Q$ ($Q>>1$), stripe domains with magnetization 
perpendicular to the surface are favored, leading to surface 
magnetic charges. Since hcp Co has an intermediate
Q value ($Q=0.35$), numerical modeling of the film micromagnetic 
structure is necessary to determine equilibrium domain
configurations.  It has been shown numerically that in hcp Co   
DWs branch, being Bloch-like in the film center and forming flux 
closure caps at the top and bottom surface of the film to reduce the 
magnetostatic energy \cite{Ebels}. 

\begin{figure}[bt]
\epsfxsize=3.5in
\centerline{\epsfbox{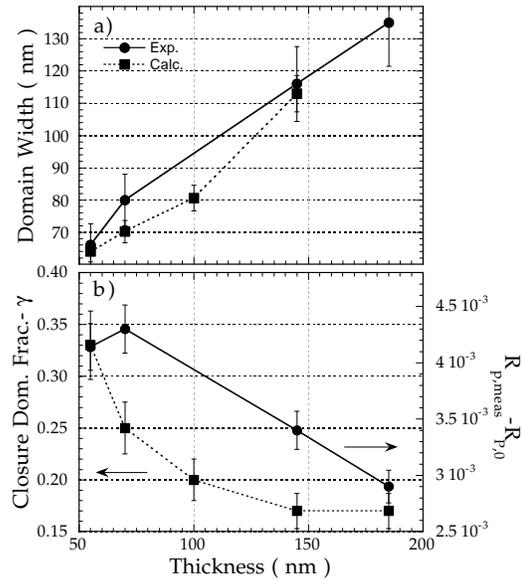}}
\vspace{-2.5 cm}
\caption{(a) Domain size versus film thickness; experimental (solid circles)
and calculated values (solid squares).
(b) The calculated in-plane magnetization 
volume (solid squares) and the magnitude of the MR 
$R_{p,meas}-R_{P,0}$ in the  perpendicular geometry
(solid circles) as a function of wire thickness.
Inset: Calculated magnetic domain cross-section of a 70 nm thick Co
element showing out-of-plane magnetized stripe domains and in-plane magnetized
flux closure caps. }
\label{fig2}
\end{figure}

The magnetic structure of films of the thicknesses studied has been computed 
in zero field with the LLG Micromagnetics Simulator \cite{Scheinfein}.
The equilibrium 
magnetization is found from the minimization of the system's free energy 
composed of exchange, magnetocrystalline anisotropy, magnetostatic, and 
Zeeman terms. The time evolution of the magnetization is given by the 
Landau-Lifschitz-Gilbert equation \cite{Aharoni}. The magnetization
distribution is approximated by a discrete cubic mesh, with a cell volume
of 1000 $nm^3$ and 
tests performed using a finer grid have shown similar results. As seen in Fig. 
2a, such calculations produce domain widths which are in fair agreement 
with experiment. The inset of Fig. 2b shows a part of the simulated magnetic 
cross section of a 70 nm thick Co element (with overall dimensions of 1500 
nm x 500 nm x 70 nm), where the arrows indicate the magnetization direction 
of the stripe and flux closure caps. Flux closure caps 
constitute approximately $25$ \% of the total wire volume, which is also an
approximate measure of the in-plane magnetized volume. For all Co 
wire thicknesses investigated the closure cap volumes (in-plane
magnetization)  were 
calculated as shown on the left-hand axis of Fig. 2b. In
increasing the wire thickness from 55 nm to 185 nm the in-plane
magnetization volume 
decreases from $33$~\% to $17$~\%.

\begin{figure}[bt]
\epsfxsize=3.25in
\centerline{\epsfbox{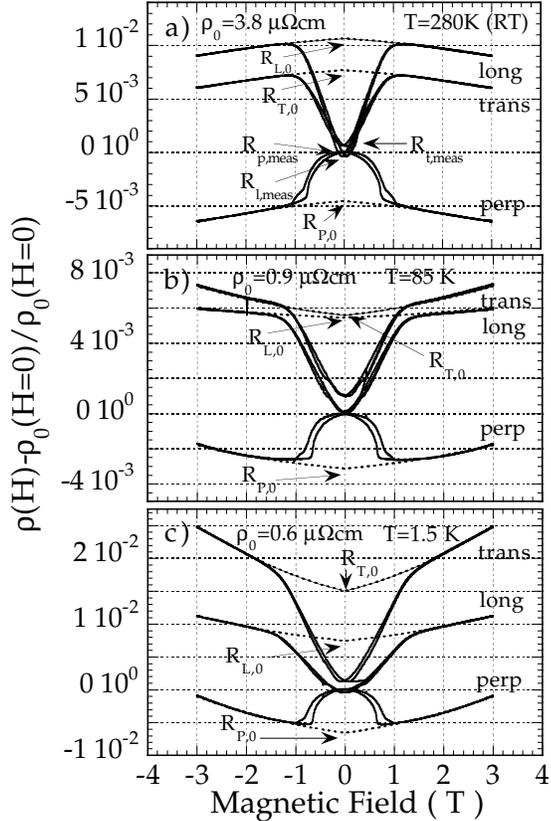}}
\caption{MR data of a 5 $\mu$m linewidth 55 nm thick Co wire in the 
perpendicular, transverse, and longitudinal field geometries at a) room 
temperature, b) 85 K, and c) 1.5 K.}
\label{fig3}
\end{figure}

MR measurements were performed using a variable temperature high 
field cryostat with in-situ rotation capabilities. The resistivity
of the Co wires was measured using a 4 probe ac ($\sim 10$ Hz) bridge
technique with currents of 10 
$\mu $A to 100 $\mu $A. The applied magnetic field was oriented
in-plane both parallel 
(longitudinal geometry) and perpendicular (transverse geometry) to the long 
wire axis (i.e., the current direction) as well as perpendicular to the
film plane 
(perpendicular geometry). Fig. 3(a) shows such measurements performed at 
room temperature on a 55 nm thick film. The low field MR is 
positive for in-plane magnetic fields 
and negative for perpendicular applied fields. Hysteresis is also evident, 
particularly in the perpendicular MR, which correlates well 
with magnetization hysteresis loops (Fig. 4). Above the saturation field
($\sim 1.4$ T) there is a large anisotropy of the resistivity, with 
the resistivity largest 
when the magnetization is in the film plane and parallel to the current. 
As generally observed in ferromagnetic materials, the 
resistivity depends on the angle of the current and magnetization as well as 
the angle the magnetization makes with respect to the crystallographic axes. 
These anisotropies have their origin in the spin-orbit interaction
and the fact 
that the orbital moment depends on the orientation of the magnetization in 
the crystal \cite{McGuire,Campbell}.

\begin{figure}[bt]
\epsfxsize=2.95in
\centerline{\epsfbox{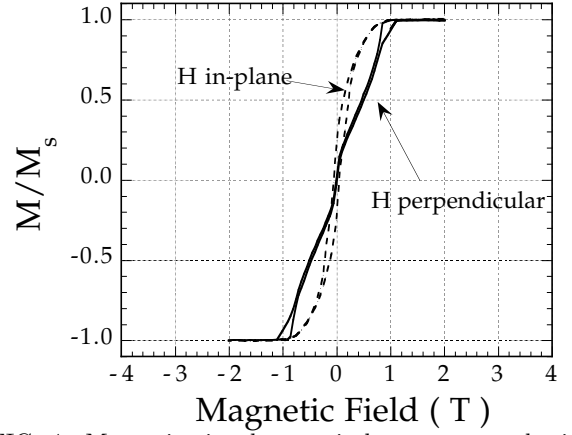}}
\caption{Magnetization hysteresis loops measured with a SQUID magnetometer of a
55 nm thick Co sample at 300 K for applied fields in-plane (dashed line) 
and perpendicular to the film plane (solid line).}
\label{fig4}
\end{figure}

This resistivity anisotropy is important in the interpretation
of the low field MR because the magnetization in zero applied field has
components along all three dimensions. For example, for the CPW geometry
(as illustrated in Fig. 1), the magnetization
of the stripe domains are out-of-the-film plane and perpendicular to the 
current, the magnetization of the flux closure caps are in-plane and 
parallel to the current, and the magnetization of the Bloch wall rotates 
through an orientation in-plane and perpendicular to current. Thus a 
saturating field will both erase DWs and reorient the magnetization 
with respect to the current and crystal. The low field MR which results 
from resistivity anisotropy and the reorientation of 
the film magnetization was neglected in the initial work on hcp Co films,
as it was incorrectly assumed that the magnetization and current 
remain always perpendicular in zero applied field \cite{Gregg}. 

This contribution can be estimated within an effective medium model 
of the resistivity. In the limit in which the electron
mean free path is smaller than the domain size \cite{Knote1} 
and the resistivity anisotropy is small, the resistivity can be written as
a weighted average of the resistivities, to first order in the resistivity
anisotropy, $\epsilon_L = R_{L,0}-R_{P,0}$ and $\epsilon_T = R_{T,0}-R_{P,0}$.
Then starting from the maze configuration (Fig. 1a) the 
perpendicular MR is :
\begin{equation}
R_{P,meas}-R_{P,0} = 
\gamma (\case{1}{2}(R_{L,0}+R_{T,0}) - R_{P,0}) + 
O(\epsilon_{L}^2,\epsilon_{T}^2)
\end{equation}
where $\gamma$ is the volume of in-plane magnetized closure caps.  Here 
$R_{L,T,P,0}$ are the MR extrapolated from high field
to H=0 (dashed lines in Fig. 3), as will be described below, and
normalized to the resistivity measured at 
H=0 in the maze configuration, $\rho_0(H=0)$ ($R_{P,meas}$
is taken to be the zero of the MR, see Fig. 3). 
In this expression, the small volume of
in-plane magnetized DW material has been neglected, only the flux closure
caps are considered. 
Within this picture, the negative MR observed in the perpendicular field 
geometry is due to the erasure of higher resisitivity closure caps in the applied 
field. Further, the magnitude of the perpendicular
MR is thickness dependent because the volume of the in-plane
magnetized material depends on sample thickness (Fig. 2). For example, 
from the MR measurements shown in Fig. 3(a) on a 55 nm thick film
and with $\gamma = 0.33$, $R_{P,0}$  is estimated to 
be $-4.5 \: \times \: 10^{-3}$, in close correspondence with the 
measured perpendicular MR. Fig. 2b shows that the perpendicular MR 
generally increases with increasing in-plane magnetized volume fractions.
The difference between CPW and CIW resistivities (i.e. 
$R_{t, meas} - R_{l, meas}$), associated with rotating the 
magnetization direction of the flux closure caps from parallel
(or antiparallel) to perpendicular to the current, 
in Fig. 3(a) is given by,
$ \gamma (R_{L,0} - R_{T,0}) = 1 \:  \times \: 10^{-3}$, in close agreement
with the experimental value.  
Such estimates show that the predominate MR effects observed in this 
material are explicable by film micromagnetic structure and resistivity 
anisotropy, without the need to invoke DW scattering effects. 

Temperature dependent resistivity measurements for CPW and CIW geometries
show more interesting behavior, which is
not explicable simply in terms of ferromagnetic resistivity anisotropy.
With decreasing temperature the in-plane resistivity
anisotropy changes sign \cite{Ulrich}, due to the increasing
importance of the anisotropy in
the Lorentz MR.  
The anisotropy of the Lorentz MR is important at low temperature
because of the large internal fields within ferromagnetic domains even in the
absence of externally applied fields. The Lorentz 
MR is larger for fields (and hence magnetization) transverse 
to the current, while spin-orbit coupling (AMR) leads
to larger in-plane resistivity for magnetization parallel to the 
current \cite{McGuire}.  
More quantitatively, the Lorentz MR
is an even function of $B/\rho \sim \omega_c\tau $, the cyclotron frequency
times the relaxation time, where $B$ is the internal field in the 
ferromagnet; $B=4\pi M+H-H_d$,
$H$ the applied field and $H_d$ the demagnetization field. With the film
magnetization oriented in-plane, the internal field for Co is $4\pi M = 1.8 \: T$.
To determine the $H=0$ resistivity anisotropy, the MR data above
the saturation field are fit to $aB^2=a(4\pi M + H - H_d)^2$, with fitting
parameter $a$.  These fits and their
extrapolation to $H=0$ are shown 
in Fig. 3. In Fig. 3(b) it is seen that the in-plane resistivity anisotropy is 
nearly zero ($R_{L,0}=R_{T,0}$) at $85 K$, 
which we denote the compensation temperature, $T_{comp}$. At lower temperature
$R_{T,0}$ is greater than $R_{L,0}$ (Fig. 3c).  

At the compensation temperature
differences in CPW and CIW resistivities due to in-plane resistivity
anisotropy should approach zero--as changing the orientation of the DWs
rotates the flux closure caps, yet will produce no change in film resistivity. 
In Fig. 3b a small difference in CPW and CIW resistivities is observed,
$\Delta_{tl} = 9 \: \times \: 10^{-4}$. Further, while the resistivity 
anisotropy changes sign and becomes larger in magnitude with decreasing
temperature (Fig. 5), $\Delta_{tl}$ is always positive.  This implies that the 
difference between CPW and CIW resistivities is not due simply to 
resistivity anisotropy, which would be proportional to $R_{L,0} - R_{T,0}$.
This is illustrated in Fig. 5 in which the difference in in-plane resistivity
anisotropy and $\Delta_{tl}$ are plotted versus temperature.

\begin{figure}[bt]
\epsfxsize=2.95in
\centerline{\epsfbox{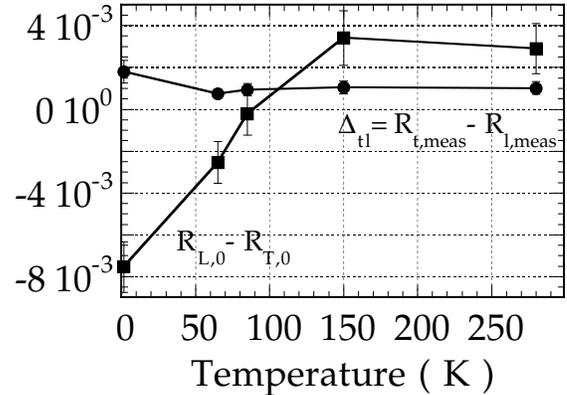}}
\caption{Temperature dependence of difference between CPW and CIW resistivities, 
$\Delta_{tl}$, and  $R_{L,0} - R_{T,0}$ of a 55 nm thick Co wire.}
\label{fig5}
\end{figure}

The greater CPW resistivity is consistent with a small additional resistivity
due to DW scattering, however there is also another possible physical explanation
for this result, which we discuss below. First, to get an idea of the the 
order of magnitude of any intrinsic DW scattering contribution to the 
resistivity, we assume that
$\Delta_{tl}$ at $T_{comp}$ is due to DW scattering.
Since walls will be much more effective at increasing
resistivity when arranged perpendicular to the current, we further assume
DWs have only a small effect on resistivity
when parallel to the current \cite{Knote2}. 
The DW interface resistivity is then given
by $r= \frac{d}{\delta} \Delta_{tl} \rho_o \delta =
\Delta_{tl} \rho_o d$,
where d is the domain size, $\delta$ is the wall width ($\sim 15 \:  nm$)
and $\rho_o$ is the film resistivity.  The Table summarizes the 
MR measurements at the compensation temperature and these estimations for
different wire thicknesses.
For the films
studied the average interface resistance is
$6 \: \pm \: 2 \: \times \: 10^{-19} \: \Omega m^2$ at $T_{comp}$ and the
MR due to the DW material,  
$\Delta \rho_{wall}/ \rho_o = \frac{d}{\delta} \Delta_{tl},$ is $0.5 \% $.
For comparison, these values are approximately a factor of 100 
smaller than the Co/Cu interface
resistance and MR in GMR multilayers with current 
perpendicular to the plane of the
layers \cite{Gijs}. 

\begin{figure}[bt]
\epsfxsize=2.95in
\centerline{\epsfbox{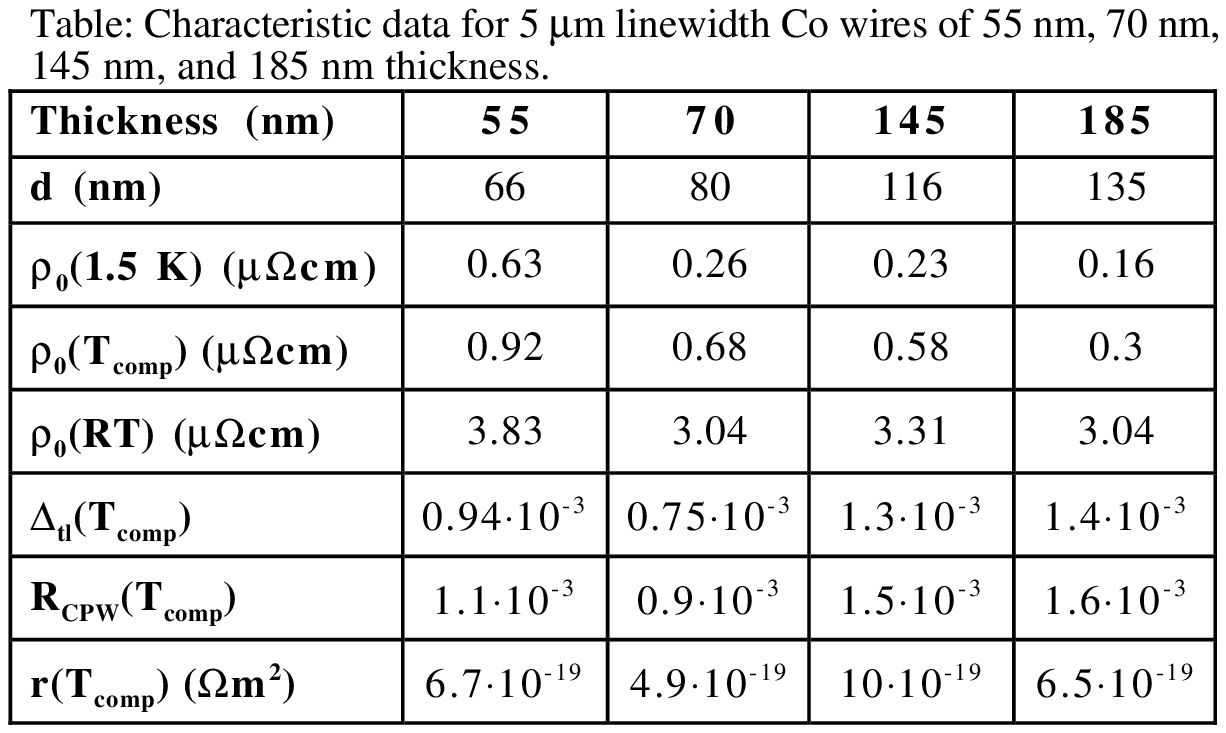}}
\end{figure}

Another mechanism which could produce the observed offset between
CPW and CIW resistivities involves the Hall effect \cite{Berger}. Both the 
ordinary Hall effect and anomalous Hall effect in ferromagnetic materials lead to an 
angle between the current and the electric field in the sample. The ordinary
Hall effect in zero applied field, is again associated with the large internal
fields within ferromagnetic domains. For the CPW geometry the electric
field will be normal to the DWs, except in a very narrow region near the sample
boundaries (within about a domain width, 100 nm). For this reason there will be 
a deflection of the current in the sample. As the
Hall angle changes sign in alternating magnetization domains, the current will
zig-zag through the sample. Berger found that this mechanism would lead to
$R_{CPW}-R_{CIW} \simeq (\omega_c\tau)^2$ \cite{Berger}.  
Interestingly, at 85 K for the 55 nm thick film
we estimate this to be $4 \times 10^{-4}$, about half the observed difference.
   
In summary, the large negative MR at room temperature for fields 
applied along the easy axis of hcp Co films with stripe domains is due 
to the film micromagnetic structure and ferromagnetic
resistivity anisotropy. The temperature dependence of the difference
between CPW and CIW resistivities shows that the intrinsic effect of
DW interface scattering is at most a small effect on the resistivity 
of such a stripe domain material.  The Hall effect may be important to
explaining the observed offset between CPW and CIW resistivities. 

\section*{Acknowledgments}
The authors thank Peter M. Levy for helpful discussions
of the work and comments on the manuscript. We thank M. Ofitserov
for technical assistance.
This research was supported by DARPA-ONR, Grant \# N00014-96-1-1207.  
Microstructures were prepared at the CNF, project \#588-96.  
\\
\\
$^*$Corresponding author: andy.kent@nyu.edu


\begin{references}
\bibitem{Knote1}
At room temperature we estimate the electron mean free path to be 30 nm.
\bibitem{Knote2}
For example, in the model of Ref. 11, a ratio of CPW to CIW MR due to
DW scattering is 7 (see their eqn. 21), for a spin assymmetry appropriate
for Co (ref. 23) 
$\rho^{\uparrow}_o/\rho^{\downarrow}_o= 3$.
\end{references}
\end{document}